\documentclass[prd,aps,showpacs,tightenlines,nofootinbib]{revtex4}
\usepackage{graphicx}

 \newcommand{\CL}{{\cal L}}
 
\newcommand{\CO}{{\cal O}}

\newcommand{\bear}{\begin{array}}  \newcommand{\eear}{\end{array}}
\newcommand{\bea}{\begin{eqnarray}}  \newcommand{\eea}{\end{eqnarray}}
\newcommand{\beq}{\begin{equation}}  \newcommand{\eeq}{\end{equation}}
\newcommand{\bef}{\begin{figure}}  \newcommand{\eef}{\end{figure}}
\newcommand{\bec}{\begin{center}}  \newcommand{\eec}{\end{center}}
\newcommand{\non}{\nonumber}  
\newcommand{\lmk}{\left(}  \newcommand{\rmk}{\right)}
\newcommand{\lkk}{\left[}  \newcommand{\rkk}{\right]}
\newcommand{\lhk}{\left \{ }  \newcommand{\rhk}{\right \} }
\newcommand{\del}{\partial}  
\newcommand{\vect}[1]{\mbox{\boldmath${#1}$}}
\newcommand{\bib}{\bibitem} 
\newcommand{\la}{\left\langle} \newcommand{\ra}{\right\rangle}

\newcommand{\vex}{\mbox{\boldmath${x}$}}
\newcommand{\mpl}{M_{Pl}}

\def\IB#1#2#3{{\bf #1}, #2 (19#3)}
\def\IBB#1#2#3{{\bf #1}, #2 (20#3)}
\def\IBID#1#2#3{{\it ibid}. {\bf #1}, #2 (19#3)}
\def\IBIDD#1#2#3{{\it ibid}. {\bf #1}, #2 (20#3)}

\def\APJLL#1#2#3{Astrophys. J. Lett. {\bf #1}, L#2 (20#3)}

\def\JP#1#2#3{J. Phys. A {\bf #1}, #2 (19#3)}

\def\MPLA#1#2#3{Mod. Phys. Lett. A {\bf #1}, #2 (19#3)}

\def\NAT#1#2#3{Nature (London) {\bf #1}, #2 (19#3)}
\def\NATT#1#2#3{Nature (London) {\bf #1}, #2 (20#3)}
\def\NPB#1#2#3{Nucl. Phys. {\bf B#1}, #2 (19#3)}

\def\PLB#1#2#3{Phys. Lett. B {\bf #1}, #2 (19#3)}

\def\PLBold#1#2#3{Phys. Lett. {\bf#1B}, #2 (19#3)}

\def\PRD#1#2#3{Phys. Rev. D {\bf #1}, #2 (19#3)}
\def\PRDD#1#2#3{Phys. Rev. D {\bf #1}, #2 (20#3)}

\def\PRL#1#2#3{Phys. Rev. Lett. {\bf#1}, #2 (19#3)}

\def\PRT#1#2#3{Phys. Rep. {\bf#1}, #2 (19#3)}

\def\PTP#1#2#3{Prog. Theor. Phys. {\bf #1}, #2 (19#3)}

\begin{document}

\title{Quantitative evolution of global strings from the Lagrangian viewpoint}
\author{Masahide Yamaguchi}
\affiliation{Physics Department, Brown University, Providence, Rhode Island
  02912 \\ and \\
Research Center for the Early Universe, University of
  Tokyo, Tokyo 113-0033, Japan}
\author{Jun'ichi Yokoyama}
\affiliation{Department of Earth and Space Science, Graduate School of
Science, Osaka University, Toyonaka 560-0043, Japan}

\date{\today}


\begin{abstract}
  We clarify the quantitative nature of the cosmological evolution of
  a global string network, that is, the energy density, peculiar
  velocity, velocity squared, Lorentz factor, formation rate of loops,
  and emission rate of Nambu-Goldstone bosons, based on a new type of
  numerical simulation of scalar fields in Eulerian meshes.  We give a
  detailed explanation of a method to extract the above-mentioned
  quantities to characterize string evolution by analyzing scalar
  fields in Eulerian meshes from a Lagrangian viewpoint.  We confirm
  our previous claim that the number of long strings per horizon
  volume in a global string network is smaller than in the case of a
  local string network by a factor of $\sim$10 even under cosmological
  situations, and its reason is clarified.
\end{abstract}

\pacs{98.80.Cq \hspace{6cm} BROWN-HET-1332, OU-TAP-188}

\maketitle


\section{Introduction}

The idea of spontaneous symmetry breaking in high energy physics has
profound implications on the vacuum structure of the Universe and our
Universe has presumably experienced various phase transitions, which
are thermal \cite{Kibble} or nonthermal \cite{KVY}. Their consequences
may be traced by topological defects that may have been produced with
them.  Indeed, many implications of topological defects for cosmology
have been investigated \cite{VS}. Furthermore, recently, defect
formations have also been discussed in the context of phase
transitions which occur in the laboratory. For example, defect
formations in $^{3}$He \cite{He3} and $^{4}$He \cite{He4} are studied
in detail. Thus the cosmological scenario of defect formation can be
tested by experiments in the laboratory \cite{Zurek}.
  
Among several types of topological defects, strings hold a unique
position in cosmology, because they do not overclose the Universe,
unlike magnetic monopoles or domain walls, settling down to a scaling
solution in which the typical scale of the network grows in proportion
to the horizon scale \cite{Kibble,Kibble2}.  The key mechanism to
achieve a scaling behavior is intercommutation of infinite strings to
dissipate their energy by producing closed loops, which subsequently
decay to radiation of relativistic particles or gravitational waves
depending on their property.  While one may understand the qualitative
nature of the scaling solution analytically, more quantitative
features such as the number of long strings per horizon volume or the
size spectrum of loops produced cannot be obtained unless full
numerical analyses are performed.

Although there exist two types of strings, namely, local and global
strings depending on the nature of the symmetry breaking, only the
former have been investigated extensively for a long time as far as
numerical analysis is concerned. By numerically solving
the equation of motion of string segments derived from the Nambu-Goto
action \cite{NG}, several groups have confirmed the scaling behavior
\cite{AT,AT2,BB,AS}, and estimated the scaling parameter $\xi$ as
$\xi\simeq 10$ in the radiation dominated universe \cite{AT2,BB,AS}.
Here, $\xi$ is defined as
\beq
  \xi = \rho_{\infty} t^{2} / \mu
  \label{eq:xi}
\eeq
where $\rho_{\infty}$ is the energy density of long strings and $\mu$
is the string tension per unit length. Thanks to this feature, local
strings generate density fluctuations with a scale-invariant spectrum
and their cosmological consequences were investigated extensively some
time ago.  Recent observations of the cosmic microwave background
anisotropy \cite{cmb}, however, disfavor the cosmic-string scenario of
structure formation, and the motivations to investigate local strings
as a source of primordial density fluctuations have somewhat
diminished, although a hybrid model of structure formation may still
be viable, where primordial fluctuations are comprised of adiabatic
fluctuations induced by inflation and isocurvature perturbations by
topological defects \cite{TDCMB}.\footnote{Note that topological
  defects can be compatible with cosmic inflation \cite{KVY}. Note
  also that local strings may still be important in that they may emit
  massive particles as sources of ultrahigh energy cosmic rays
  \cite{VHS}.}

On the other hand, global strings are much better motivated in the
context of axion cosmology. They are formed as a consequence of the
breaking of the Peccei-Quinn U(1) symmetry \cite{PQ,VE}, which was
introduced to solve the strong $CP$ problem in quantum chromodynamics.
These global strings radiate axions as associated Nambu-Goldstone (NG)
bosons \cite{Davis}, which are one of the most promising candidates
for cold dark matter. Despite their importance, the cosmological
evolution of global strings has been less studied and the results of
local strings have often been borrowed even though there is a decisive
difference between them. For local strings, the gradient energy of
scalar fields is canceled out by gauge fields far from the core. The
string core is well localized and the false vacuum energy of the core
dominates the system. Hence the Nambu-Goto action is suitable as an
effective action in order to study the cosmological evolution of local
strings except at crossing \cite{NG}. On the other hand, for global
strings, there are no gauge fields to cancel the gradient energy of
the NG scalar field, which dominates over the false vacuum energy of
the core. The effective action appropriate for global strings is not
the Nambu-Goto action but the Kalb-Ramond action, which incorporates
NG bosons and their couplings with the core \cite{KR}. Also, due to
the gradient energy of the NG scalar field, a long-range force works
between global strings. Thus, the behavior of the two types of string
is expected to be different and it is nontrivial whether global
strings relax to a scaling regime.  In fact, for example, the behavior
of global monopoles is quite different from that of local monopoles
due to long-range forces. While the former may be useful in cosmology
\cite{monopole,Yamaguchi}, the latter causes a disaster
\cite{Preskill} unless diluted by inflation.  Furthermore, it has been
shown in the literature that local and global strings behave
differently in two dimensional space \cite{YB,2dim}.

There have been several attempts to investigate the cosmological
evolution of the global string network by use of the Kalb-Ramond
action \cite{BS,sharp}. However, the Kalb-Ramond action is too
complicated to be dealt with numerically. It has difficulty because of
logarithmic divergence due to the self-energy of the string. In such a
situation the authors and Kawasaki made the first numerical
investigations of the cosmological evolution of global strings without
resort to the Kalb-Ramond action. We instead solved the equations of
motion for scalar fields forming strings in three dimensional Eulerian
meshes \cite{YKY}.  We found that the global string network would also
go into a scaling regime but the scaling parameter $\xi$ was found to
be of the order of unity \cite{YKY,YYK}, which is significantly
smaller than the case of local strings ($\xi \simeq 10$).

Recently, however, this quantitative difference was questioned and it
was claimed that the smallness of the scaling parameter of global
strings might be a fake due to the small dynamic range of the
numerical simulations \cite{MS,MSM}. The authors of \cite{MS,MSM}
claimed that in our simulations global strings lost their energy by
excessive direct emission of NG bosons.  Based on the speculation that
such direct emission of NG bosons from long strings would be
negligible on cosmological scales, they reached a conjecture that both
types of strings behave quantitatively in the same way in the
cosmological context, namely, $\xi \simeq 10$, with the only
difference between them being the energy loss mechanism of closed
loops.

In order to clarify the evolution of global strings on cosmological
scales, we should first study which is the dominant energy loss
mechanism of long strings in numerical simulations, loop production or
direct emission of NG bosons. We note that the master equation for
the energy density of long strings, $\rho_{\infty}$, can be expressed
as
\beq 
\frac{d\rho_{\infty}}{dt} = - 2 H (1 + \la v^{2} \ra)
\rho_{\infty} - \Gamma_{\rm loop}\rho_{\infty} - \Gamma_{\rm
  NG}\rho_{\infty},
  \label{eq:energyloss}
\eeq
where the second and the third terms on the right-hand side represent
energy loss due to loop formation and direct emission of NG bosons or
axions, respectively, and $\la v^{2} \ra$ denotes the average square
velocity of string segments.  For our purpose, we need to calculate
both the loop production rate $\Gamma_{\rm loop}$ and the emission
rate of NG bosons $\Gamma_{\rm NG}$, and compare their magnitudes in
simulations.

If the system relaxes to the scaling regime, which will be confirmed
to be the case shortly, the string network is described by the
so-called one-scale model\footnote{Although improvements of the
  simplest one-scale model have been proposed by taking into account
  effects of small-scale structures \cite{ACK} and time evolution of
  the velocity \cite{MS,MS2}, the original one-scale model is
  sufficient here because small-scale structures are smeared in the
  case of global strings and the velocity remains constant in the
  scaling regime, as will be shown later. } with the characteristic
scale $L \equiv \sqrt{\mu / \rho_{\infty}}$, which grows with the
horizon scale $L \propto t$.  If we introduce the loop production
coefficient $c$ and the emission coefficient $\kappa$ of the NG bosons
by
\bea
\Gamma_{\rm loop}\rho_{\infty}= c \frac{\rho_{\infty}}{L},~~~~~
  \Gamma_{\rm NG}\rho_{\infty} 
     = \kappa \frac{\rho_{\infty}}{L}, 
  \label{eq:ckappa}
\eea
these parameters remain constant in the one-scale picture and are
related to the scaling parameter $\xi$ as
\beq
  \xi = \lmk \frac{1 - \la v^{2} \ra}{c + \kappa} \rmk^{2}.
  \label{eq:relation}
\eeq
Indeed, if $\kappa$ incorrectly turned out to be much larger than $c$,
we would find a smaller value of $\xi$ than it should really be.
Hence, it is essential to evaluate the parameters $c$ and $\kappa$ in
the scaling regime.

In our previous simulations \cite{YKY}, however, it was impossible to
calculate these quantities for two reasons. First, in the previous
work a lattice was identified as part of a string based on the value
of the potential energy there, which had the microscopic problem that
we occasionally found disconnected string pieces, although the overall
features were traced reasonably well. Second, it was impossible to
monitor intercommutation of strings for lack of dynamical information,
namely, the velocity of each string segment.  These problems are
overcome by our new identification scheme and Lagrangian analysis of
the evolution of strings \cite{YY}.

The rest of the paper is organized as follows. In Sec. II we present
the method of our new procedure to follow the Lagrangian evolution of
global strings in Eulerian simulations, that is, new methods of
identification of strings, measurement of string velocity, and
estimation of the intercommutation rate. In Sec. III the results are
described and applied to cosmological situations. Finally, Sec. IV is
devoted to a discussion and conclusion.

\section{numerical simulations}

\subsection{Formulation}
\label{sub:formulation}

We consider the following Lagrangian density for two-component real
scalar fields $\phi_{a}(x)$ ($a = 1, 2$) which can accommodate global
strings:
\beq
  \CL[\phi_{a}] = \frac12 g_{\mu\nu}
                   (\del^{\mu}\phi_{a})(\del^{\nu}\phi_{a})
                    - V[\phi_{a},T],
  \label{eq:lagrangian}
\eeq 
in the spatially flat Robertson-Walker spacetime,
\beq
 ds^2=g_{\mu\nu}dx^{\mu}dx^{\nu}=dt^2-R^2(t)d\vect{x}^2,
\eeq
with $R(t)$ being the scale factor.  We adopt the following potential
at finite temperature $T$:
\bea
  V[\phi_{a},T] &=& \frac{\lambda}{4}(\phi^{2} - \sigma^2)^2 
                 + \frac{\lambda}{6}T^2\phi^{2} \\
                &=& \frac{\lambda}{4}(\phi^{2} - \eta^2)^2
                 +  \frac{\lambda}{4}(\sigma^{4} - \eta^4).
  \label{eq:potential}
\eea
Here $\phi^{2} \equiv \phi_{1}^{2} + \phi_{2}^{2}$ and $\eta^{2}
\equiv \sigma^{2} - T^{2}/3 = \sigma^{2}(1 - T^{2}/T_{c}^{2})$ with
$T_{c} \equiv \sqrt{3}\sigma$ being the critical temperature. Below
the critical temperature, global U(1) symmetry is broken to form
global strings.

The equations of motion for the scalar fields are given by
\beq
  \ddot{\phi_{a}}(x) + 3H(t)\dot{\phi_{a}}(x) 
    - \frac{1}{R(t)^2}\nabla^2\phi_{a}(x)
      + \frac{\del V}{\del \phi_{a}} = 0,
  \label{eq:EOM}
\eeq
where an overdot represents the time derivative. The discretization of
these differential equations is given in Appendix \ref{app:1}. Our
numerical calculations are based on the staggered leapfrog method with
second order accuracy both in time and in space. In the radiation
dominated Universe, the Hubble parameter $H(t) = \dot R(t)/R(t)$ and
cosmic time $t$ are given by
\bea
  H(t)^2 = \frac{8\pi}{3 \mpl^2} \frac{\pi^2}{30} g_{*} T^4,
   ~~~~~
  t = \frac{1}{2H} \equiv \frac{\epsilon}{T^2},
  \label{eq:hubble}
\eea
where $\mpl = 1.2 \times 10^{19}$ GeV is the Plank mass and $g_{*}$ is
the total number of degrees of freedom for the relativistic particles.
We define a dimensionless parameter $\zeta$ as
\beq
  \zeta \equiv \frac{\epsilon}{\sigma}  = \lmk \frac{45\mpl^2}
      {16\pi^3g_{*} \sigma^2}
  \rmk^{1/2},
  \label{eqn:zeta}
\eeq
and take $\zeta=10$ and the self-coupling $\lambda = 0.08$ for
definiteness, but these particular choices do not affect the results.
We start the numerical simulation at the temperature $T_{i} = 2 T_{c}$
corresponding to $t_{i} = t_{c}/4$ and adopt as an initial condition
the thermal equilibrium state with a mass equal to the inverse
curvature of the potential at that time.  In this state $\phi_a$ and
$\dot{\phi}_a$ are Gaussian distributed with the correlation
functions
\bea
  \la \beta|\phi_a(\vect x)\phi_b(\vect y)|\beta
                             \ra_{\rm equal~time} &=&
   \int \frac{d^3 k}{(2\pi)^3} \frac1{2\sqrt{\vect k^2 + m^2}}
           \coth{\frac{\beta\sqrt{\vect k^2 + m^2}}{2}}
             e^{i\vect k \cdot (\vect x-\vect y)}\delta_{ab} \:,   \\
             && \non \\
  \langle \beta|\dot\phi_a(\vect x)\dot\phi_b(\vect y)|\beta
                             \rangle_{\rm equal~time} &=&
   \int \frac{d^3 k}{(2\pi)^3} \frac{\sqrt{\vect k^2 + m^2}}{2}
           \coth{\frac{\beta\sqrt{\vect k^2 + m^2}}{2}}
             e^{i\vect k \cdot (\vect x-\vect y)}\delta_{ab} \:, 
\eea
where $m^2=V''[\phi_a,T_i]$ and $\beta=T^{-1}_i$.  $\phi_a(\vect x)$
and $\dot\phi_a(\vect y)$ are uncorrelated for $\vect x \ne \vect y$.
We generate the initial configuration in the momentum space, where the
scalar fields are uncorrelated. Then they are transformed into the
position space by fast Fourier transformation.

We perform simulations in five different sets of lattice sizes and
spacings as shown in Table \ref{tab:set1} to investigate their effects
on the results. In all cases, the time step is taken as $\Delta t =
0.01 t_{i}$, and the periodic boundary condition  is adopted. In
the case (a), the box size is nearly equal to the horizon length
$H^{-1}$ and the lattice spacing to the typical core width $\delta \sim
1.0/(\sqrt{2\lambda}\sigma)$ of a string at the final time $t_{f} =
200 t_i$. The other cases have equal or larger simulation volumes.

In this type of numerical calculation, it is a nontrivial task to
identify string cores from the data of scalar fields because a point
with $\phi_{a} = 0$ corresponding to a string core is not necessarily
situated at a lattice point. So, in the next subsection, we give our
new method of identification of string cores.

\subsection{Identification method of strings}
\label{sub:identification}

In previous work \cite{YKY,YYK}, whether a lattice point was a part of
string core was judged based on the potential energy density there.
That is, a lattice point was identified to be at the core of a string
if the potential energy density there was found to be larger than that
of a static cylindrically symmetric solution of a global string at
$r=\Delta x_{\rm phys}/\sqrt{2}$ off center, where $\Delta x_{\rm
  phys}$ is the physical length of the lattice spacing at each time.

Although this method worked fairly well to evaluate the scaling
parameter, it was inadequate to fully identify strings particularly
for small loops which do not resemble the static cylindrically
symmetric solution due to the curvature.  As a result we occasionally
found disconnected string segments that should not exist in reality.
Furthermore, it was impossible to find a more correct position of the
string core in a box beyond the lattice spacing, which is extremely
important to evaluate the length and velocity of the string correctly.

\begin{figure}[htb]
\begin{center}
\begin{minipage}{80mm}
\includegraphics[width=8cm]{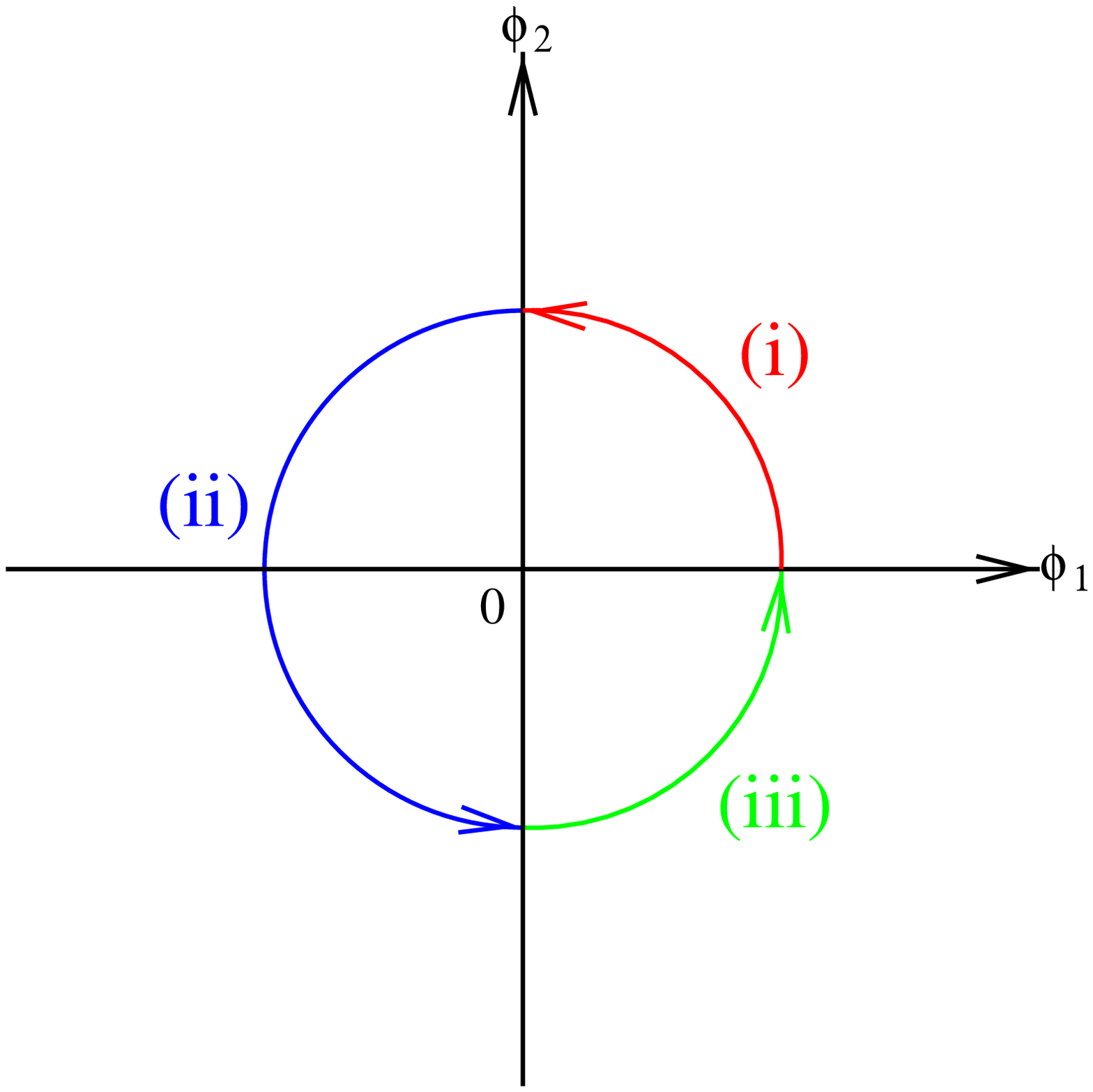}
\begin{center} 
\label{fig:phase1}
\end{center}
\end{minipage}
\hspace{1cm}
\begin{minipage}{80mm}
\includegraphics[width=8cm]{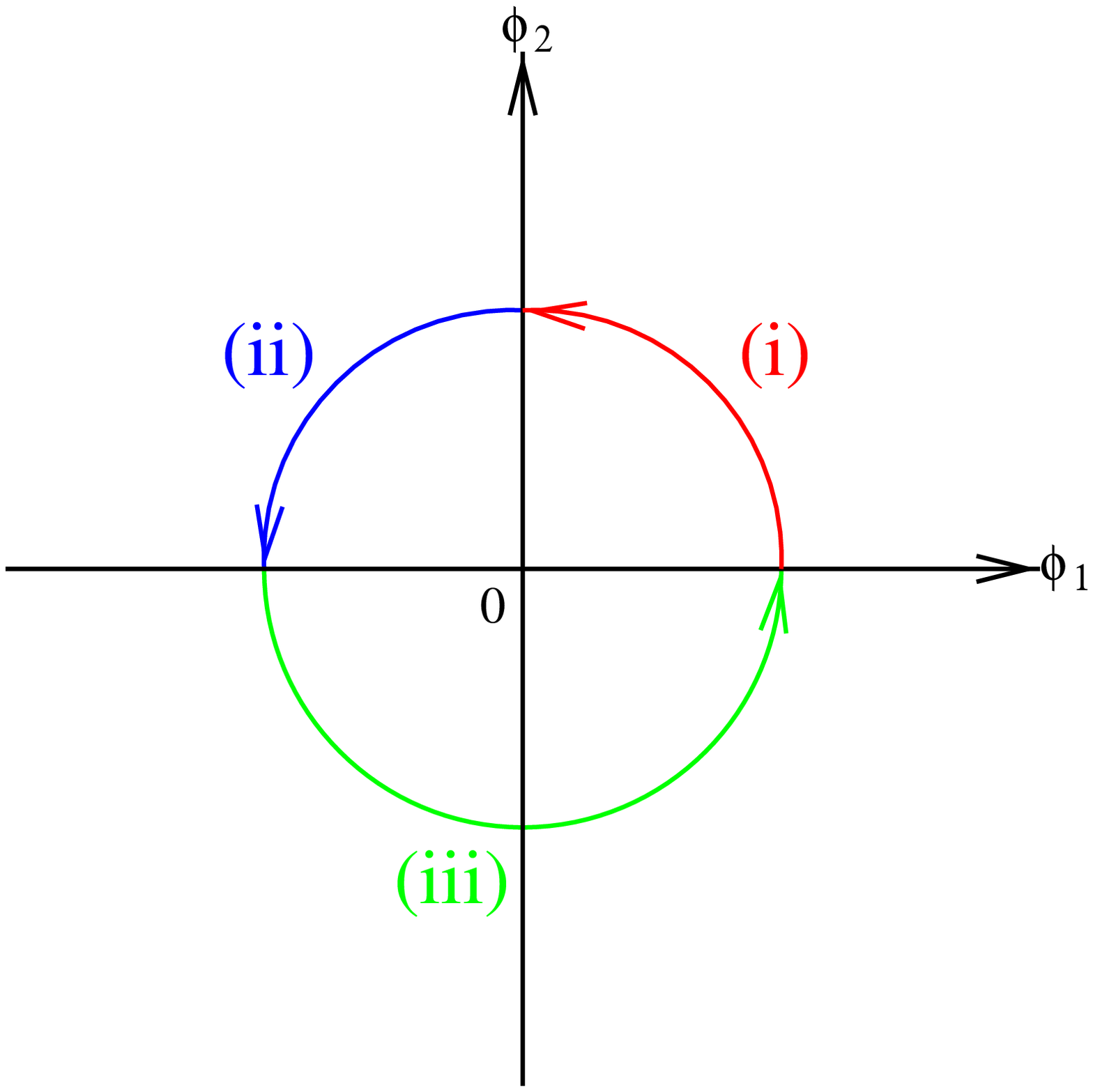}
\begin{center} 
\label{fig:phase2}
\end{center}
\end{minipage}
\caption{Left: 
  Relative phase of the scalar fields is classified into three groups,
  (i) $0 \le \theta < \frac{\pi}{2}$, (ii) $\frac{\pi}{2} \le \theta <
  \frac{3\pi}{2}$, (iii)
  $\frac{3\pi}{2} \le \theta < 2\pi$.\\
  Right: Another classification of the relative phase, (i) $0 \le
  \theta < \frac{\pi}{2}$, (ii) $\frac{\pi}{2}\le\theta < \pi$, (iii)
  $\pi \le\theta < 2\pi$, was also tried but the numerical results did
  not depend on these choices.}
\label{fig:phase}
\end{center}
\end{figure}

Here we develop a new method of string identification based on the
fact that strings lie on the intersection of two surfaces
$\phi_1(\vect{x},t)=0$ and $\phi_2(\vect{x},t)=0$, so that if a string
penetrates a (sufficiently small) plaquette, $\phi_a$ has a different
sign at one or two corners of the plaquette from the rest for each
$a$.  First, we classify the relative phase of the scalar fields into
three groups as shown in Fig.\,\ref{fig:phase} (left), that is,
\bea
   \theta \equiv \arccos\frac{\phi_{2}}{\sqrt{\phi_{1}^2+\phi_2^2}}
 +2\pi \lkk 1-\Theta (\phi_2) \rkk, 
\eea \bea
  {\rm (i)}   &&\quad 0 \le \theta < \frac{\pi}{2}, \non \\
  {\rm (ii)}  &&\quad \frac{\pi}{2} \le \theta < \frac{3\pi}{2},  \\
  {\rm (iii)} &&\quad \frac{3\pi}{2} \le \theta < 2\pi, \non
\eea
where the arccosine should take the principal value and $\Theta
(\phi_2)$ is the step function.

\begin{figure}[htb]
\includegraphics[width=6cm]{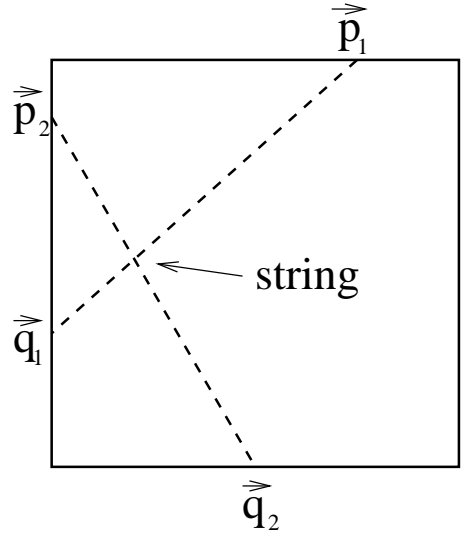} 
\caption{$\vect p_{a}$ and $\vect q_{a}$ are points corresponding to
  $\phi_{a} = 0$ for each $a$.  These points are obtained by linear
  interpolation using the values of $\phi_a$ at two corners of the
  plaquette between which it changes sign.  The line $\phi_{a} = 0$
  for each $a$ is drawn by simply connecting $\vect p_{a}$ and $\vect
  q_{a}$ by a straight line.  The intersection of these two lines is
  identified as the position through which a string penetrates a
  plaquette.}
\label{fig:cross1}
\end{figure}

Then, we judge whether a string penetrates a plaquette by monitoring
the phase rotation around it just as in the Vachaspati-Vilenkin
algorithm \cite{VV}. If we assigned an equal range $2\pi/3$ of the
relative phase to all regions (i)-(iii) as in the original
Vachaspati-Vilenkin algorithm and judged the presence of a string from
the phase rotation, we would occasionally identify a plaquette as
containing a string even if $\phi_1$ or $\phi_2$ takes the same sign
at its four corners.  This is the main reason why all regions
(i)-(iii) do not have equal ranges of the relative phase $\theta$ in
our scheme, in which this case is avoided and $\phi_a$ takes a
different sign for each $a$ at one or two corners of each plaquette
along which a circular phase rotation is observed. Then, using the
values of $\phi_a$ at its four corners we can draw two lines
corresponding to $\phi_1=0$ and $\phi_2=0$ and their intersection is
identified with the point where a string penetrates the plaquette, as
shown in Fig.\,\ref{fig:cross1}.

\begin{figure}[htb]
\includegraphics[width=7.7cm]{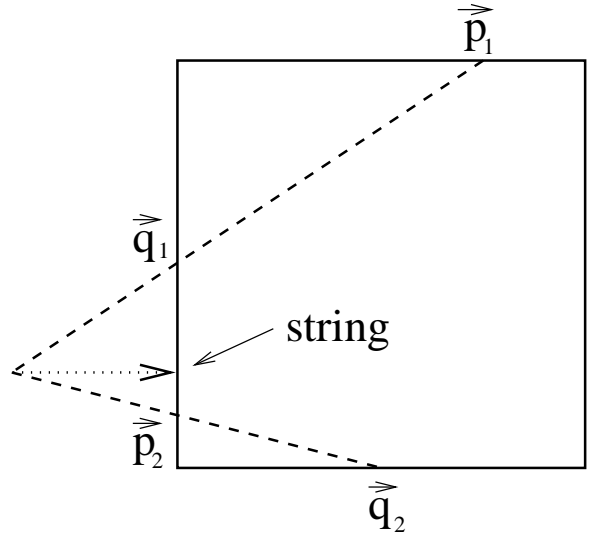} 
\caption{When the intersection is found outside the plaquette, the
  nearest point on the edge of the plaquette is identified as the
  penetration point of the string.}
\label{fig:cross2}
\end{figure}

In fact, the intersection point could be found outside the plaquette.
In this case, the nearest point on the edge of the plaquette from the
intersection is identified as the position of the string as shown in
Fig.\,\ref{fig:cross2}.  In our simulations we encountered such a case
very rarely and it did not cause any serious problems.

By joining these points, the strings are completely connected and a
more accurate total length of the global string can be obtained.  One
may wonder if our biased classification of the relative phase may
cause some artificial effects on the result of numerical simulations.
So we have tried another classification of the relative phase as shown
in Fig.\,\ref{fig:phase} (right), that is,
\bea
  {\rm (i)}   &&\quad 0 \le \theta < \frac{\pi}{2}, \non \\
  {\rm (ii)}  &&\quad \frac{\pi}{2} \le \theta \le \pi, \quad  \\
  {\rm (iii)} &&\quad \pi \le \theta < 2\pi , \non
\eea
and confirmed that the results do not depend on the classification
scheme for the relative phase.

At each time step, we can evaluate the total length of the global
string by the method shown above.  Since strings typically move at a
speed close to unity, we should multiply the length of each string
segment by $\mu = \gamma \mu_{s}$ to calculate the total energy
density of strings. Here $\gamma$ is the Lorentz factor and
$\mu_{s}\simeq 2\pi\eta^2\ln\lkk t/(\delta\xi^{1/2})\rkk$ is the line
density of a static string \cite{MSM}. So it is important to establish
a method to calculate the velocity and Lorentz factor of the string
segments, which will be presented shortly.  We can then evaluate the
scaling parameter $\xi$ from Eq.\,(\ref{eq:xi}).

\subsection{Velocity of strings}
\label{sub:velocity}

Here we describe our method to evaluate the velocity of strings, which
is a nontrivial task in Eulerian calculations of scalar field
configurations.\footnote{The measurement of the velocity was also
  attempted in Ref. \cite{MSM}. In their method, the velocity is
  obtained by comparing the positions of strings at different times.
  Although some efforts have been made to reduce errors, their method
  may cause some systematic errors because the time separation cannot
  be shorter than that sufficient to ensure a string can move to
  another lattice. On the other hand, our method is much superior; it
  is based on first principles and the velocity of each string segment
  is given by a local quantity.}

First we expand the scalar fields $\phi_{a}(\vect x,t_0+\delta t)$
around $\phi_{a}(\vect x_{0},t_{0})$ up to first order,
\bea
  \phi_{a}(\vect x,t_0+\delta t) \cong \phi_{a}(\vect x_{0},t_{0}) 
      + \nabla\phi_{a}(\vect x_{0},t_{0}) \cdot (\vect x-\vect x_{0}) 
       + \dot\phi_{a}(\vect x_{0},t_{0})\delta t
      \qquad ( a = 1, 2).
  \label{eq:expansion}
\eea
Suppose that a string core exists at a point $\vect x_0$ at time $t_0$
and moves to a point $\vect x$ at $t=t_0 + \delta t$, that is,
$\phi_a(\vex_0,t_0)=0$ and $\phi_a(\vex,t_0+\delta t)=0$ for each $a$.
Then, from Eq. (\ref{eq:expansion}) we find that the loci of the
string core at $t=t_0+\delta t$ lie on the intersection of two planes,
\beq
 \nabla\phi_{a}(\vect x_{0},t_{0}) \cdot (\vect x-\vect x_{0})
 +\dot\phi_{a}(\vect x_{0},t_{0})\delta t=0,  \label{plane}
\eeq
with $a=1,2$. Since motion tangential to a string is a gauge mode, we
should evaluate the velocity normal to it. Suppose that the line normal to
the string segment at $(\vect x_0,t_0)$ reaches across the above
intersection line at $\vect x =\vect x_{l}(\vect x_0,t_0,\delta t)$.
Then we can easily obtain the velocity of this string segment as
\bea
  \vect{v}(\vect x_0,t_0) &=& \lim_{\delta t \longrightarrow 0}
\frac{\vect x_{l}(\vect x_0,t_0,\delta t) - \vect x_{0}}{\delta t}
\nonumber \\
    &=& \left.\frac{\dot\phi_{1}\nabla\phi_{2} - \dot\phi_{2}\nabla\phi_{1}}
           {|\nabla\phi_{1} \times \nabla\phi_{2}|}\right|_{\vect{x_0},t_0}.
  \label{eq:velocity}
\eea    
We calculate the velocity at each point where the strings cross
plaquettes.  For this purpose we need to evaluate $\dot\phi_{a}$ and
$\nabla\phi_{a}$ at arbitrary points on a plaquette, as shown
explicitly in Appendix \ref{app:2}, where $\dot\phi_{a}$ and
$\nabla\phi_{a}$ are given by the quantities on lattice points up to
the second order.  Collecting the value of Eq. (\ref{eq:velocity}) at
each intersection of strings with plaquettes, we can obtain the
average of the velocity, the velocity squared, and the Lorentz factor.

\subsection{Intercommutation of strings}
\label{sub:intercommutation}

Now we discuss the energy loss rate from infinite strings to loops by
calculating the intercommutation rate $c$.  In simulations of cosmic
strings based on the Nambu-Goto action, one has to assign the
intercommutation probability of intersecting strings by hand due to
the lack of microscopic information. Since the strings are identified
from the scalar field configuration in our simulation, we can
unambiguously calculate how strings intersect with each other, namely,
whether they simply pass through or intercommute upon collision.  We
can therefore identify new loops by monitoring the lengths of all long
strings and loops at each time step and comparing the data with those
at the previous time step.

\subsection{NG boson emission}
\label{sub:emission}

It is very difficult to directly evaluate the NG boson emission rate
because separation of the emitted axions and NG phase associated with
the string core is nontrivial. Since loops also emit these particles,
it is a formidable task to identify axions radiated from long strings
only.  However, we have already described how we can obtain
$\rho_{\infty}$, $\langle v^2 \rangle$, and $\Gamma_{\rm loop}$ from
simulation data at each time step, and $d\rho_{\infty}/dt$ can also be
calculated from $\rho_{\infty}$ at two adjacent time steps.  So all
the quantities in the master equation (\ref{eq:energyloss}) can be
calculated from our simulation data except for the last term, which
can be found from the equation itself.  In particular, in the scaling
regime, when the one-scale description suffices, we can find $\kappa$
from \beq \kappa = \frac{1 - \la v^{2} \ra}{\sqrt{\xi}} - c .
  \label{eq:relationk}
\eeq

\section{Results}

\subsection{Scaling parameter}

\begin{figure}[htb]
\includegraphics[width=18cm]{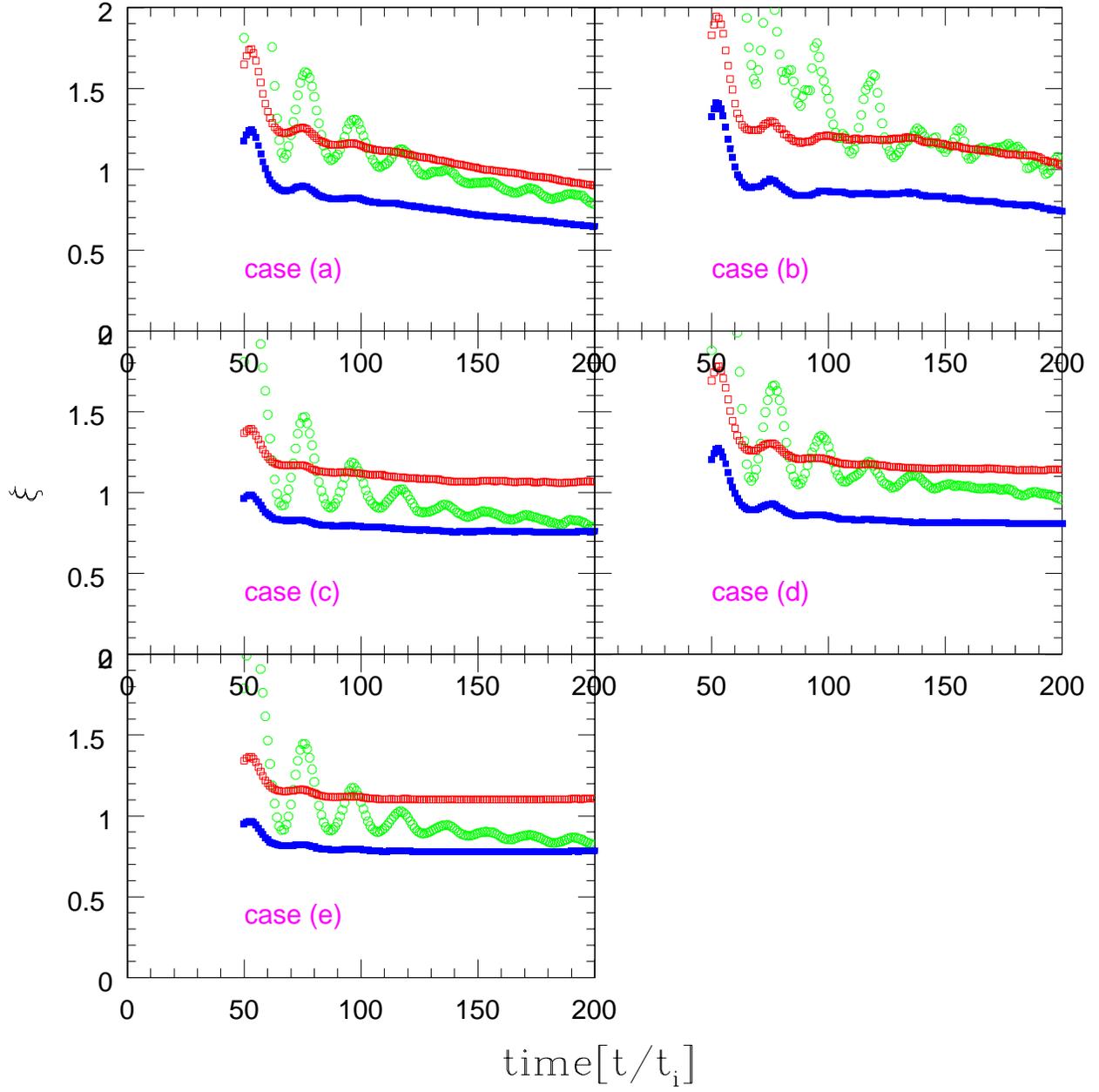} 
\caption{Time evolution of the scaling parameter $\xi$. 
  Filled squares represent our new identification method. Blank
  circles correspond to our previous identification method based on
  the potential energy density \cite{YKY,YYK}. Blank squares are
  results of the Vachaspati-Vilenkin algorithm.}
\label{fig:xi}
\end{figure}

In Fig.\,\ref{fig:xi}, the time evolution of $\xi$ is depicted for all
cases [(a)-(e)]. The filled squares represent the time development of
$\xi$ for the new identification method.  For comparison, the results
of our previous identification method based on the potential energy
density \cite{YKY,YYK}, which slightly overestimates $\xi$ by a factor
of $1.2$, are shown by blank circles. The results of the original
Vachaspati-Vilenkin identification scheme are indicated by blank
squares, obtained by counting the number of boxes through which a
string passes as was done in Ref. \cite{MSM}. This method
overestimates $\xi$ by a factor of $1.4$.

We find that differences in the simulation settings except the box
size normalized by the final horizon scale do not affect the overall
behavior significantly.  $\xi$ becomes smaller in the smallest box
size with $= H^{-1}$ at the final time. This is mainly because under
the periodic boundary conditions, a string feels an attractive force
from the boundary and tends to disappear. Indeed, in the case (a)
this artificial effect starts to operate before the system really
relaxes to the scaling solution, so $\xi$ continues decreasing without
a plateau.  On the other hand, in the case (b), the boundary effect
becomes operative after $t\simeq 130$ when $\xi$ starts to decrease
from the scaling value. The other three cases, whose box size is
larger than $2 H^{-1}$, are free from the boundary effects and the
scaling parameter $\xi$ remains constant.  Hence we conclude that
after the relaxation period the global string network goes into the
scaling regime. From Table \ref{tab:set2}, which shows average values
of $\xi$ after some relaxation period ($t > 80$), $\xi$ is found to be
$\xi \simeq 0.80$ in the scaling regime.

\subsection{String velocity}

\begin{figure}[htb]
\includegraphics[width=18cm]{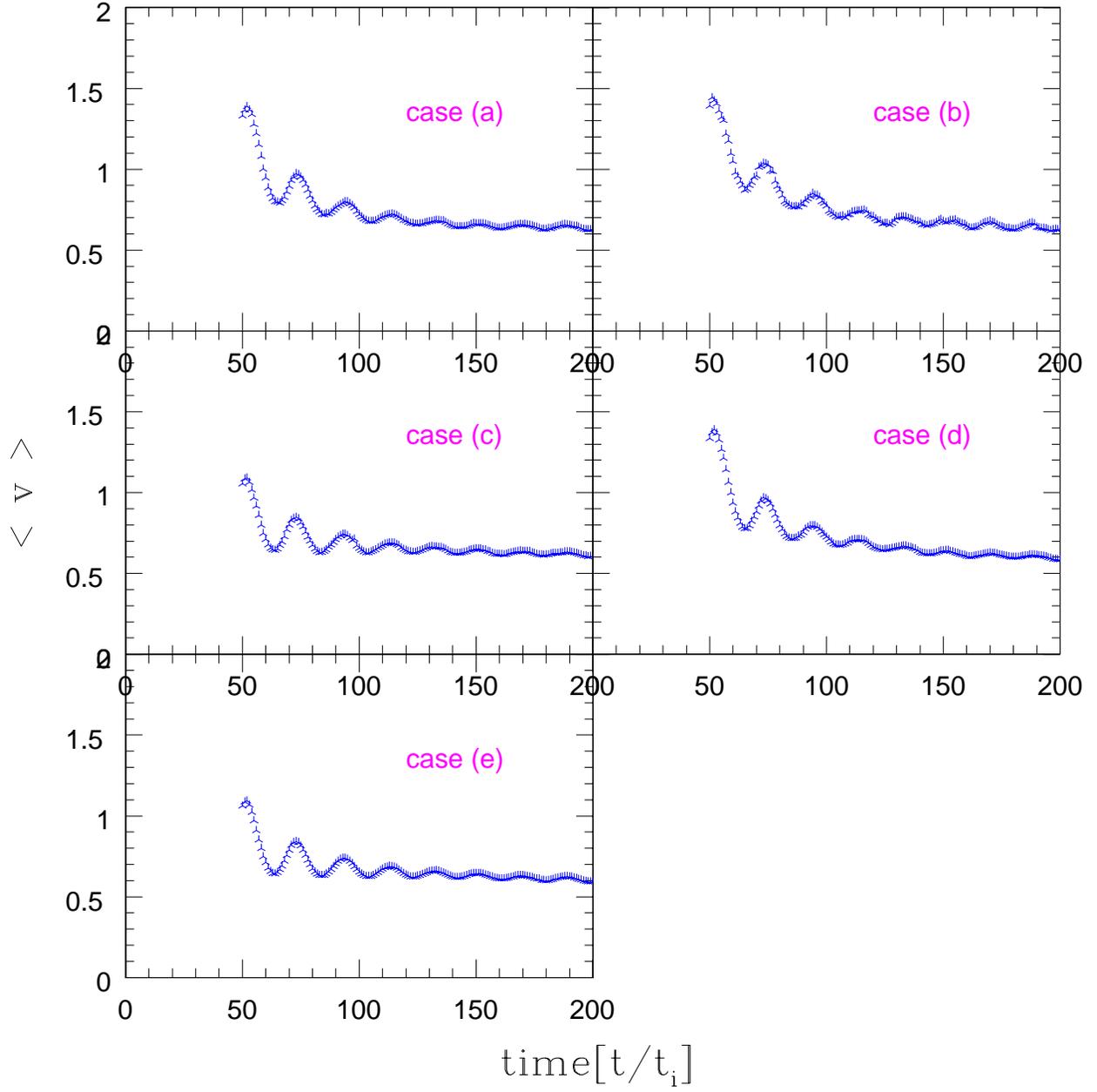} 
\caption{Time evolution of average velocity of global strings is
shown.}
\label{fig:velocity}
\end{figure}

\begin{figure}[htb]
\includegraphics[width=18cm]{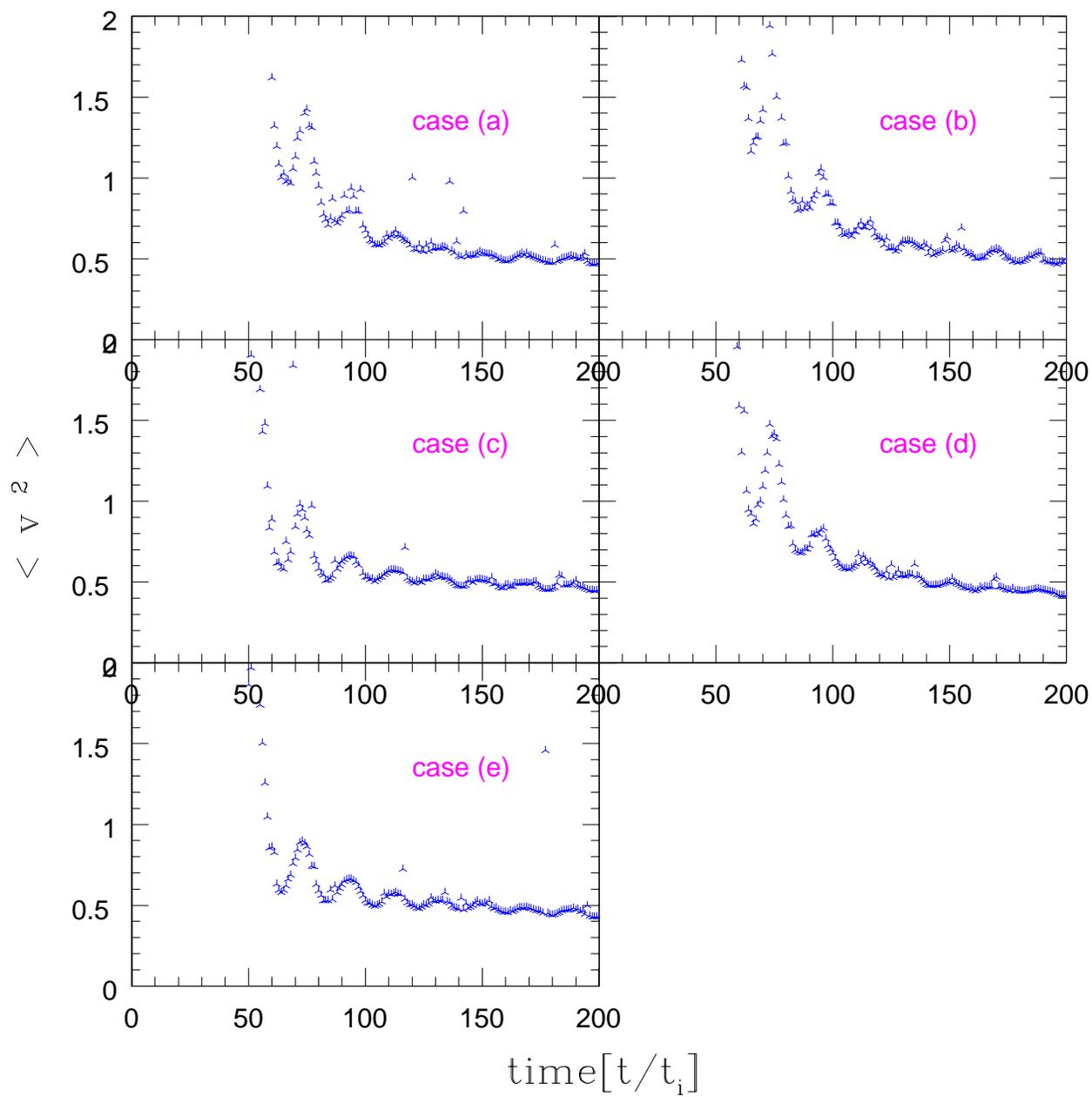} 
\caption{Time evolution of average square velocity.}  
\label{fig:svelocity}
\end{figure}

\begin{figure}[htb]
\includegraphics[width=18cm]{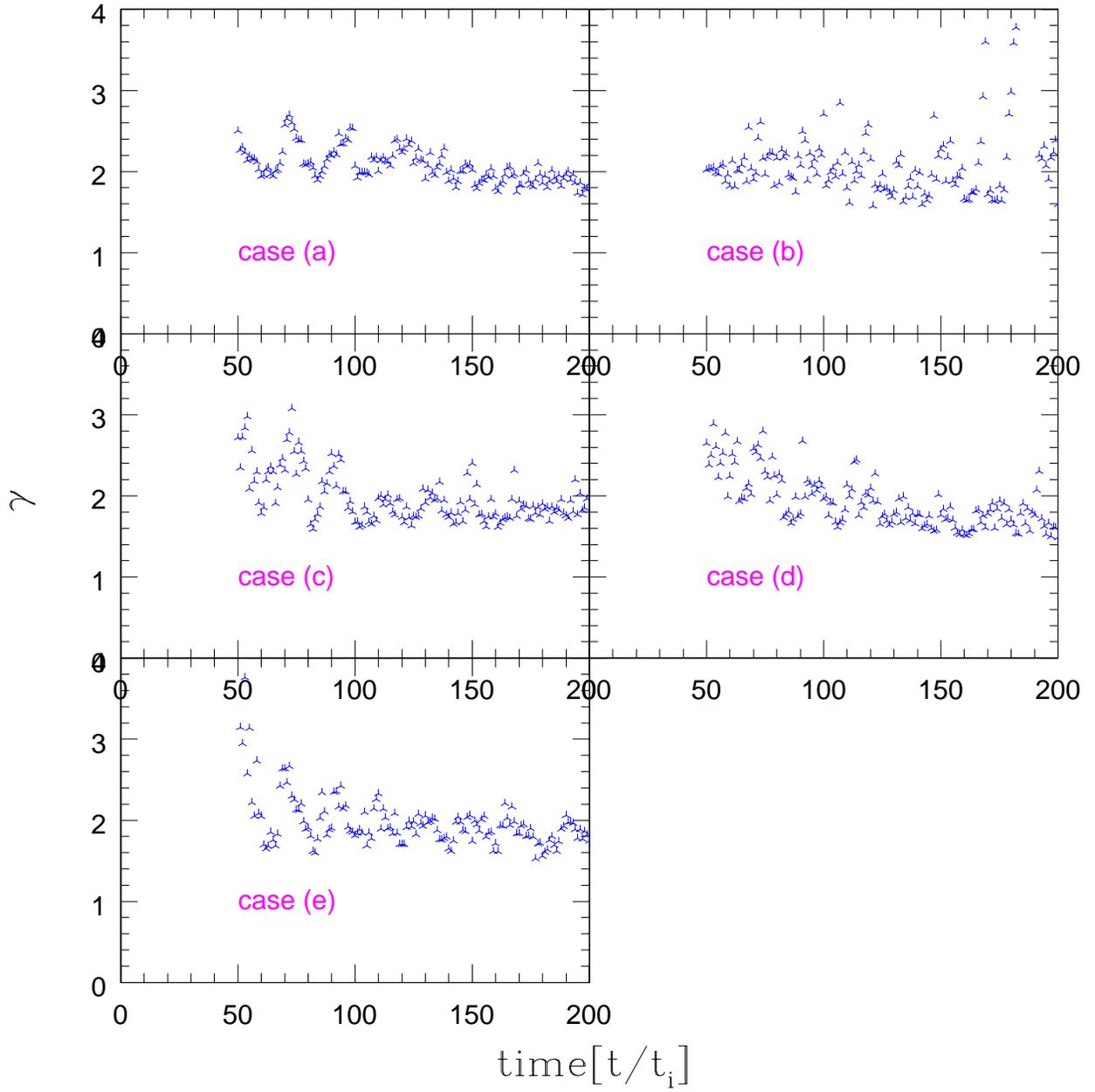} 
\caption{Time evolution of average Lorentz factor.}  
\label{fig:gamma}
\end{figure}

In Figs. \ref{fig:velocity} -- \ref{fig:gamma}, the time evolutions of
the average velocity $\la v\ra$, the average square velocity $\la
v^2\ra$, and the Lorentz factor $\la \gamma \ra$ are shown. After some
relaxation period, they become fairly constant in all cases.  The
average values are given in Table \ref{tab:set2}. We find that the
average values become larger in the smallest box size. This is mainly
because under the periodic boundary conditions, a string feels an
attractive force from the boundary and tends to be accelerated.
However, such an effect becomes negligible for a box size larger than
$2 H^{-1}$ as observed in Table \ref{tab:set2}. Thus we find the
average velocity $\la v \ra \simeq 0.60$ and the average square
velocity $\la v^{2} \ra \simeq 0.50 \gg \la v \ra^{2}$, in the scaling
regime.  On the other hand, the average Lorentz factor has a large
scatter in time, although the long-time average is fairly constant
with $\langle\bar{\gamma}\rangle \simeq 1.8$. Since string segments
moving with a speed close to the light velocity have extremely large
Lorentz factors and push up the average dramatically, the fluctuation
in the number of such string segments results in this large scatter.
This is the reason the average Lorentz factor $\langle\gamma\rangle
=1.8\pm 0.2$ is much larger than $1/\sqrt{1 - \la v^{2} \ra}$ and $
1/\sqrt{1 - \la v \ra^{2}}$.  Thus the energy per unit length of a
string is enhanced by a factor $\langle\bar{\gamma}\rangle \simeq 1.8$
compared with a static string.

\subsection{Energy dissipation coefficients}

As explained in Sec. \ref{sub:intercommutation} we can obtain the
intercommutation rate $c$ without assigning its probability by hand.
The result is given in Table \ref{tab:set2}, where we find that $c$ is
larger in the smallest box size. This can be understood in the same
way, namely, under the periodic boundary conditions, a string feels an
attractive force from the boundary and tends to intercommute more
often. Since such an effect is unimportant for a box size larger
than $2 H^{-1}$ as discussed above, $c$ is found to be $c = 0.40 \pm
0.04$.

Taking account of the relation (\ref{eq:relationk}), the emission
parameter $\kappa$ is given in Table \ref{tab:set2}. In the
simulations with the smallest box, strings intercommute more often due
to the boundary effect, which suppresses NG boson emission. But again
such an effect is inoperative for cases with box size larger than
$2H^{-1}$.  So we conclude that the NG boson emission rate is $\kappa
= 0.16 \pm 0.04$.

\section{Discussion and conclusion}

In this paper we have investigated the cosmological evolution of the
global string network in detail. In our numerical simulations, the
equations of motion of the two-component real scalar field are solved
on Eulerian meshes. In order to follow the time evolution of global
strings, we have developed a new identification method which enables
us to find a more correct position of the string core in a box beyond
the lattice spacing. Furthermore, we have given a detailed explanation
of the method to extract Lagrangian quantities, such as velocity and
intercommutation, characterizing the evolution of global strings. The
NG boson emission rate is obtained from the master equation
(\ref{eq:energyloss}). Thus the quantitative nature of the
cosmological evolution of the global string network is elucidated
without setting the intercommutation probability of two intersecting
strings by hand.  Specifically, we find the scaling parameter
characterizing the energy density $\xi \simeq 0.80$, the peculiar
velocity $\la v \ra \simeq 0.60$, the velocity squared $\la v^2 \ra
\simeq 0.50$, the Lorentz factor $\langle\gamma\rangle = 1.8 \pm 0.2$,
the formation rate of loops $c = 0.40 \pm 0.04$, and the emission rate
of Nambu-Goldstone bosons $\kappa = 0.16 \pm 0.04$ from our results.

In our simulations, due to the limit of the dynamic range of the
lattices, the logarithm of the ratio of the Hubble radius to the
string width, which appears in the expression for the effective line
density of global strings, took $\ln(t/\delta)\sim 5$, while in the
cosmological setting it can be as large as $\CO (100)$.  The authors
of \cite{MS,MSM} argue that $\kappa$ is inversely proportional to
$\ln(t/\delta)$, although $c$ and $\la v^2 \ra$ do not have such
dependence. Based on this speculation, they claimed that we obtained a
smaller value of the scaling parameter $\xi$ in our previous work
\cite{YKY} than it should really be, because $\kappa$ was incorrectly
large there. They also argued that in the cosmological situation loop
production is the dominant mechanism of energy dissipation of long
strings and that $\xi$ would take the same value as for local strings.

Now that we have all the values of relevant quantities from our
simulation data, we can disprove their claim. Indeed, we find that
$\kappa$ is smaller than $c$ and hence direct emission of NG bosons
does not dominate the energy loss mechanism even in our setting of
numerical simulations with a relatively small dynamic range,
$\ln(t/\delta)$. If we extrapolate the value of $\kappa$ with the
scaling $\kappa \propto 1/\ln(t/\delta)$, we find $\kappa \lesssim
0.01$ in the cosmological situation. Even then the scaling parameter
(\ref{eq:relation}) does not increase much, yielding $\xi = 1.6 \pm
0.3$, which is still much smaller than that of local strings.

Thus we conclude that there is a quantitative difference between the
cosmological evolution of global strings and that of local strings.
It is based on the qualitative difference that global strings have a
long-range force and intercommute more often with larger $\la v^2 \ra$
than do local strings.  Note that it is by no means surprising that
global defects behave differently from local defects in cosmology.  In
the case of monopoles, as discussed in the Introduction, global
monopoles evolve in a strikingly different manner than do magnetic
monopoles. In the case of strings, both local and global defects relax
to a scaling solution and their difference is more subtle.  So it is
not until our numerical analysis of the evolution of the latter from
the Lagrangian point of view is performed that their difference is
fully elucidated.

Finally, we use our results to obtain a constraint on the symmetry
breaking scale $\eta\equiv f_a$ of the Peccei-Quinn U(1) symmetry.
From $\xi=1.6\pm0.3$ and $\la \gamma\ra=1.8\pm0.2$, we find $f_a
\lesssim (0.16 - 1.2)\times 10^{12}$ GeV for the normalized Hubble
parameter $h=0.7$ \cite{YKY}.  We also note that the Lagrangian method
developed in this paper is directly applicable to other species of
extended objects, for instance, topological defects such as local
strings and nontopological solitons such as Q balls as well.

\section*{Note added in proof}

After we submitted the original manuscript, we became aware of an
analytic estimate of $\xi$ which reports that it is $\sim 1$ even in
the case of the local string network. We are grateful to M. Hindmarsh
for informing us of his result \cite{Hindmarsh}.

\section*{Acknowledgments}

M.Y.\ is grateful to Robert Brandenberger for his hospitality at Brown
University, where the final part of the work was done.  J.Y.\ would
like to thank Toru Tsuribe for useful comments on numerical analysis.
This work was partially supported by JSPS Grants-in-Aid for Scientific
Research, No.\ 12-08555 (M.Y.) and No.\ 13640285 (J.Y.).

\appendix

\section{Discretization of differential equations}
\label{app:1}

The equation of motion of the scalar fields is given by
\beq
  \ddot{\phi_{a}}(x) + 3H(t)\dot{\phi_{a}}(x) 
    - \frac{1}{R(t)^2}\nabla^2\phi_{a}(x)
      + \frac{\del V}{\del \phi_{a}} = 0.
\eeq
In order to discretize the above equation, it is reduced to first-order
differential equations,
\bea
  \dot{\phi}_{a} &\equiv& \pi_{a} \non \\
  \dot{\pi}_{a} &=& -3H(t)\pi_{a}
        - \frac{1}{R(t)^2}\nabla^2\phi_{a}(x) 
        - \frac{\del V}{\del \phi_{a}}. \non \\
\eea
Expanding $\phi_{a}(t,\vect x)$ and $\phi_{a}(t+\Delta t,\vect x)$
around the intermediate time step $t+\frac12 \Delta t$, we find
\beq
  \phi_{a}(t+\Delta t,\vect x) - \phi_{a}(t,\vect x) =
      \Delta t\,\dot{\phi}_{a} \lmk t+\frac12 \Delta t,\vect x \rmk
      + \CO\bigl((\Delta t)^3\bigr),               
\eeq
up to the third order.
Thus $\pi_{a}$ at the intermediate time step is given by
\beq
  \pi_{a} \lmk t+\frac12 \Delta t,\vect x \rmk = 
     \frac{\phi_{a}(t+\Delta t,\vect x) - \phi_{a}(t,\vect x)}{\Delta t}
      + \CO\bigl((\Delta t)^2\bigr).
\eeq
In the same way, $\dot{\pi}_{a}$ and $\pi_{a}$ at the time step $t$
are represented by the quantities at the two adjacent intermediate
steps $t\pm\frac12 \Delta t$,
\bea
  \dot{\pi}_{a}(t,\vect x) &=&
     \frac{\pi_{a}(t+\frac12\Delta t,\vect x) 
          -\pi_{a}(t-\frac12\Delta t,\vect x)}
          {\Delta t}
      + \CO\bigl((\Delta t)^2\bigr), \non \\
  \pi_{a}(t,\vect x) &=&
     \frac{\pi_{a}(t+\frac12\Delta t,\vect x) 
          +\pi_{a}(t-\frac12\Delta t,\vect x)}
          {2} 
      + \CO\bigl((\Delta t)^2\bigr). \label{a6}
\eea
The second-order derivatives are approximated up to the second order in 
$\Delta$ as 
\bea
  \frac{\del^2 \phi_{a}(t,\vect x)}{\del x^2} &=&
    \frac{\phi_{a}(t,x+\Delta,y,z)
         -2\phi_{a}(t,x,y,z)
         +\phi_{a}(t,x-\Delta,y,z)}
         {\Delta^2}
    + \CO(\Delta^2), \non \\
  \frac{\del^2 \phi_{a}(t,\vect x)}{\del y^2} &=&
    \frac{\phi_{a}(t,x,y+\Delta,z)
         -2\phi_{a}(t,x,y,z)
         +\phi_{a}(t,x,y-\Delta,z)}
         {\Delta^2}
    + \CO(\Delta^2), \non \\
  \frac{\del^2 \phi_{a}(t,\vect x)}{\del z^2} &=&
    \frac{\phi_{a}(t,x,y,z+\Delta)
         -2\phi_{a}(t,x,y,z)
         +\phi_{a}(t,x,y,z-\Delta)}
         {\Delta^2}
    + \CO(\Delta^2).
\eea 

Thus the fundamental equations are discretized up to the second order
in both space and time,
\bea
  && \hspace{-1.0cm} 
  \ddot{\phi_{a}}(x) + 3H(t)\dot{\phi_{a}}(x) 
    - \frac{1}{R(t)^2}\nabla^2\phi_{a}(x)
      + \frac{\del V[\phi_{a}(x)]}{\del \phi_{a}}
  \non \\
  &&
  =
  \frac{\dot{\phi}_{a}(t+\frac12\Delta t,\vect x) 
       -\dot{\phi}_{a}(t-\frac12\Delta t,\vect x)}{\Delta t}    
  + 3H(t)\frac{\dot{\phi}_{a}(t+\frac12\Delta t,\vect x) 
               +\dot{\phi}_{a}(t-\frac12\Delta t,\vect x)}
              {2} 
  \non \\
  && \qquad
  - \frac{1}{R(t)^2} 
      \sum_{l = x,y,z} \biggl\{ \frac{ 
                                \phi_{a}(t,\vect x+\vect \Delta_{l})
                                -2\phi_{a}(t,\vect x)
                                + \phi_{a}(t,\vect x-\vect \Delta_{l}) 
                                       }
                                       {\Delta^2}   
                         \biggr\}
  + \frac{\del V[\phi_{a}(t,\vect x)]}{\del \phi_{a}} 
  + \CO\bigl((\Delta t)^2,\Delta^2 \bigr) \non \\
  && = 0,
\eea
where $\Delta_{x}=(\Delta,0,0),\Delta_{y}=(0,\Delta,0)$, and
$\Delta_{z}=(0,0,\Delta)$.

In summary, in our numerical simulations
\bea
  \dot{\phi}_{a} \lmk t+\frac12\Delta t,\vect x \rmk
    &=& \frac{1}{1+3H(t)\Delta t/2}
      \Biggl[
        \lmk 1-\frac{3H(t)\Delta t}{2} \rmk 
          \dot{\phi}_{a} \lmk t-\frac12\Delta t,\vect x \rmk \non \\
    && \quad    
        + \Delta t
            \lmk 
              \frac{1}{R(t)^2} 
                \sum_{l = x,y,z} \biggl\{ \frac{ 
                                \phi_{a}(t,\vect x+\vect \Delta_{l})
                                -2\phi_{a}(t,\vect x)
                                + \phi_{a}(t,\vect x-\vect \Delta_{l}) 
                                       }
                                       {\Delta^2}   
                         \biggr\}
              - \frac{\del V[\phi_{a}(t,\vect x)]}{\del \phi_{a}}  
            \rmk
      \Biggr], \non \\
  \phi_{a}(t+\Delta t,\vect x) &=& 
      \phi_{a}(t,\vect x) 
     + \Delta t\,\dot{\phi}_{a} \lmk t+\frac12\Delta t,\vect x \rmk.     
\eea
The value of $\dot{\phi}_{a}$ at the time step $t$, which is required
to calculate the string velocity, is evaluated from Eq. (\ref{a6}).
 
\section{Quantities at an arbitrary point on a plaquette}
\label{app:2}

In order to evaluate the velocity of a string correctly, we need to
evaluate quantities at an arbitrary point on a plaquette.  That is,
$\dot{\phi}_{a}$ and $\nabla{\phi}_{a}$ within a plaquette should be
expressed by their values at its four corners. As an example, let us
consider a plaquette parallel to the $z$ plane with four corners
$(x,y,z)$, $(x+\Delta,y,z)$, $(x,y+\Delta,z)$, and
$(x+\Delta,y+\Delta,z)$ and express
$\dot{\phi}_{a}(t,x+\alpha\Delta,y+\beta\Delta,z)$ and
$\nabla\phi_{a}(t,x+\alpha\Delta,y+\beta\Delta,z)$, which we denote
collectively by $\Pi(x+\alpha\Delta,y+\beta\Delta)$ below, in terms of
their values at these four points.  Here $0 \le \alpha \le 1$ and $0
\le \beta \le 1$.

We express $\Pi$ at the four corners by an expansion around
$(x+\alpha\Delta,y+\beta\Delta)$ as
\bea
  \Pi(x,y) 
           &=& \Pi(x+\alpha\Delta,y+\beta\Delta) 
               - \alpha\Delta\frac{\del\Pi}{\del x}
               - \beta\Delta\frac{\del\Pi}{\del y} \non \\
           &&  + \frac12\alpha^2\Delta^2\frac{\del^2\Pi}{\del x^2}
               + \alpha\beta\Delta^2\frac{\del^2\Pi}{\del x\del y}
               + \frac12\beta^2\Delta^2\frac{\del^2\Pi}{\del y^2}
               + \CO(\Delta^3), \non \\
  \Pi(x+\Delta,y) 
           &=& \Pi(x+\alpha\Delta,y+\beta\Delta) 
               + (1-\alpha)\Delta\frac{\del\Pi}{\del x}
               - \beta\Delta\frac{\del\Pi}{\del y} \non \\
           &&  + \frac12(1-\alpha)^2\Delta^2\frac{\del^2\Pi}{\del x^2}
               - (1-\alpha)\beta\Delta^2\frac{\del^2\Pi}{\del x\del y}
               + \frac12\beta^2\Delta^2\frac{\del^2\Pi}{\del y^2}
               + \CO(\Delta^3), \non \\
  \Pi(x,y+\Delta) 
           &=& \Pi(x+\alpha\Delta,y+\beta\Delta) 
               - \alpha\Delta\frac{\del\Pi}{\del x}
               + (1-\beta)\Delta\frac{\del\Pi}{\del y} \non \\
           &&  + \frac12\alpha^2\Delta^2\frac{\del^2\Pi}{\del x^2}
               - \alpha(1-\beta)\Delta^2\frac{\del^2\Pi}{\del x\del y}
               + \frac12(1-\beta)^2\Delta^2\frac{\del^2\Pi}{\del y^2}
               + \CO(\Delta^3), \non \\
  \Pi(x+\Delta,y+\Delta) 
           &=& \Pi(x+\alpha\Delta,y+\beta\Delta) 
               + (1-\alpha)\Delta\frac{\del\Pi}{\del x}
               + (1-\beta)\Delta\frac{\del\Pi}{\del y} \non \\
           &&  + \frac12(1-\alpha)^2\Delta^2\frac{\del^2\Pi}{\del x^2}
               + (1-\alpha)(1-\beta)\Delta^2\frac{\del^2\Pi}{\del x\del y}
               + \frac12(1-\beta)^2\Delta^2\frac{\del^2\Pi}{\del y^2}
               + \CO(\Delta^3).
\eea

Making an appropriate combination, $\Pi(x+\alpha\Delta,y+\beta\Delta)$
can be expressed by its values at the four vertices in the plaquette,
\bea
  && \hspace{-1.5cm} 
  (1-\alpha)(1-\beta)\,\Pi(x,y) + \alpha(1-\beta)\,\Pi(x+\Delta,y)
    + (1-\alpha)\beta\,\Pi(x,y+\Delta)+\alpha\beta\,\Pi(x+\Delta,y+\Delta)
           \non \\
  && = \Pi(x+\alpha\Delta,y+\beta\Delta)
    +\frac12\alpha(1-\alpha)\Delta^2\frac{\del^2\Pi}{\del x^2}
    +\frac12\beta(1-\beta)\Delta^2\frac{\del^2\Pi}{\del y^2}
    +\CO(\Delta^3) \label{eq:third} \\
  && = \Pi(x+\alpha\Delta,y+\beta\Delta) + \CO(\Delta^2).
\eea
In particular, replacing $\Pi(x+\alpha\Delta,y+\beta\Delta)$ by
$\dot{\phi}_{a}(x+\alpha\Delta,y+\beta\Delta)$, it can be expressed by
the quantities on the lattice points up to second order,
\bea
  \dot{\phi}_{a}(t,x+\alpha\Delta,y+\beta\Delta,z)
   &=& (1-\alpha)(1-\beta)\,\dot{\phi}_{a}(t,x,y,z)
      + \alpha(1-\beta)\,\dot{\phi}_{a}(t,x+\Delta,y,z) \non \\
   && \quad
      + (1-\alpha)\beta\,\dot{\phi}_{a}(t,x,y+\Delta,z)
      + \alpha\beta\,\dot{\phi}_{a}(t,x+\Delta,y+\Delta,z)
      + \CO(\Delta^2) .  
\eea
In the same way, inserting $\del\phi_{a}/\del x$ into $\Pi$, we find
\bea
  \frac{\phi_{a}(t,x+\alpha\Delta,y+\beta\Delta,z)}{\del x}
   &=& (1-\alpha)(1-\beta)\,\frac{\partial\phi_{a}(t,x,y,z)}{\del x}
      + \alpha(1-\beta)\,\frac{\partial\phi_{a}(t,x+\Delta,y,z)}{\del x} 
          \non \\
   && \quad
      + (1-\alpha)\beta\,\frac{\partial\phi_{a}(t,x,y+\Delta,z)}{\del x}
      + \alpha\beta\,\frac{\partial\phi_{a}(t,x+\Delta,y+\Delta,z)}{\del x}
      + \CO(\Delta^2) \\  
   &=& \frac{1}{2\Delta} \biggl[
     (1-\alpha)(1-\beta)\,
       \Bigl\{
         \phi_{a}(t,x+\Delta,y,z)-\phi_{a}(t,x-\Delta,y,z) 
       \Bigr\}
          \non \\
   && \quad \qquad
      + \alpha(1-\beta)\,
       \Bigl\{
         \phi_{a}(t,x+2\Delta,y,z)-\phi_{a}(t,x,y,z) 
       \Bigr\}
          \non \\
   && \quad \qquad
      + (1-\alpha)\beta\,
       \Bigl\{
         \phi_{a}(t,x+\Delta,y+\Delta,z)-\phi_{a}(t,x-\Delta,y+\Delta,z) 
       \Bigr\}
          \non \\
   && \quad \qquad
      + \alpha\beta\,
       \Bigl\{
         \phi_{a}(t,x+2\Delta,y+\Delta,z)
        -\phi_{a}(t,x,y+\Delta,z) 
       \Bigr\}
       \biggr]
      + \CO(\Delta^2).
\eea
Here we have used the relations
\bea
  \frac{\phi_{a}(x+\Delta,y,z)-\phi_{a}(x-\Delta,y,z)}{2\Delta}
   &=& \frac{\del\phi_{a}(x,y,z)}{\del x} + \CO(\Delta^2), \non \\
  \frac{\phi_{a}(x+2\Delta,y,z)-\phi_{a}(x,y,z)}{2\Delta}
   &=& \frac{\del\phi_{a}(x+\Delta,y,z)}{\del x} + \CO(\Delta^2).
\eea

Interchanging $x$ for $y$ and $\alpha$ for $\beta$, $\del\phi_{a}/\del
y$ is also given by
\bea
  \frac{\phi_{a}(t,x+\alpha\Delta,y+\beta\Delta,z)}{\del y}
   &=& (1-\alpha)(1-\beta)\,\frac{\partial\phi_{a}(t,x,y,z)}{\del y}
      + \alpha(1-\beta)\,\frac{\partial\phi_{a}(t,x+\Delta,y,z)}{\del y} 
          \non \\
   && \quad
      + (1-\alpha)\beta\,\frac{\partial\phi_{a}(t,x,y+\Delta,z)}{\del y}
      + \alpha\beta\,\frac{\partial\phi_{a}(t,x+\Delta,y+\Delta,z)}{\del y}
      + \CO(\Delta^2) \\  
   &=& \frac{1}{2\Delta} \biggl[
     (1-\alpha)(1-\beta)\,
       \Bigl\{
         \phi_{a}(t,x,y+\Delta,z)-\phi_{a}(t,x,y-\Delta,z) 
       \Bigr\}
          \non \\
   && \quad \qquad
      + \alpha(1-\beta)\,
       \Bigl\{
         \phi_{a}(t,x+\Delta,y+\Delta,z)-\phi_{a}(t,x+\Delta,y-\Delta,z) 
       \Bigr\}
          \non \\
   && \quad \qquad
      + (1-\alpha)\beta\,
       \Bigl\{
         \phi_{a}(t,x,y+2\Delta,z)-\phi_{a}(t,x,y,z) 
       \Bigr\}
          \non \\
   && \quad \qquad
      + \alpha\beta\,
       \Bigl\{
         \phi_{a}(t,x+\Delta,y+2\Delta,z)
        -\phi_{a}(t,x+\Delta,y,z) 
       \Bigr\}
       \biggr]
      + \CO(\Delta^2).
\eea

Using the relations obtained above and Eq.\,(\ref{eq:third}),
$\del\phi_{a}/\del z$ is calculated as
\bea
  \frac{\phi_{a}(t,x+\alpha\Delta,y+\beta\Delta,z)}{\del z}
   &=& \frac{\phi_{a}(t,x+\alpha\Delta,y+\beta\Delta,z+\Delta)
              -\phi_{a}(t,x+\alpha\Delta,y+\beta\Delta,z-\Delta)}
             {2\Delta} + \CO(\Delta^2) 
          \non \\
   && = \frac{1}{2\Delta} \biggl[
     (1-\alpha)(1-\beta)\,
       \Bigl\{
         \phi_{a}(t,x,y,z+\Delta)-\phi_{a}(t,x,y,z-\Delta) 
       \Bigr\}
          \non \\
   && \quad \qquad
      + \alpha(1-\beta)\,
       \Bigl\{
         \phi_{a}(t,x+\Delta,y,z+\Delta)-\phi_{a}(t,x+\Delta,y,z-\Delta) 
       \Bigr\}
          \non \\
   && \quad \qquad
      + (1-\alpha)\beta\,
       \Bigl\{
         \phi_{a}(t,x,y+\Delta,z+\Delta)-\phi_{a}(t,x,y+\Delta,z-\Delta) 
       \Bigr\}
          \non \\
   && \quad \qquad
      + \alpha\beta\,
       \Bigl\{
         \phi_{a}(t,x+\Delta,y+\Delta,z+\Delta)
        -\phi_{a}(t,x+\Delta,y+\Delta,z-\Delta) 
       \Bigr\}
       \biggr]
          \non \\
   && \quad
      - \frac14\alpha(1-\alpha) \Delta
       \lhk   
         \frac{\del^2 \phi_{a}(t,x+\alpha\Delta,y+\beta\Delta,z+\Delta)}
              {\del x^2} -                 
         \frac{\del^2 \phi_{a}(t,x+\alpha\Delta,y+\beta\Delta,z-\Delta)}
              {\del x^2}           
       \rhk  
          \non \\
   && \quad
      - \frac14\beta(1-\beta) \Delta
       \lhk   
         \frac{\del^2 \phi_{a}(t,x+\alpha\Delta,y+\beta\Delta,z+\Delta)}
              {\del y^2} -                 
         \frac{\del^2 \phi_{a}(t,x+\alpha\Delta,y+\beta\Delta,z-\Delta)}
              {\del y^2}           
       \rhk  
          \non \\
   && \quad
      + \CO(\Delta^2).
\eea
Although these approximations appear of first-order accuracy, using
the relations,
\bea
  && \hspace{-4.0cm} 
  \frac{\del^2 \phi_{a}(t,x+\alpha\Delta,y+\beta\Delta,z+\Delta)}
       {\del x^2} -                
  \frac{\del^2 \phi_{a}(t,x+\alpha\Delta,y+\beta\Delta,z-\Delta)}
       {\del x^2}       
     \non \\  
  && \qquad =
  2\Delta  
    \frac{\del^3 \phi_{a}(t,x+\alpha\Delta,y+\beta\Delta,z)}
         {\del x^2\del z} + \CO(\Delta^3),                  
    \non \\
  && \hspace{-4.0cm} 
  \frac{\del^2 \phi_{a}(t,x+\alpha\Delta,y+\beta\Delta,z+\Delta)}
       {\del y^2} -                
  \frac{\del^2 \phi_{a}(t,x+\alpha\Delta,y+\beta\Delta,z-\Delta)}
       {\del y^2}
    \non \\     
  && \qquad =
  2\Delta  
    \frac{\del^3 \phi_{a}(t,x+\alpha\Delta,y+\beta\Delta,z)}
         {\del y^2\del z} + \CO(\Delta^3),                 
\eea
we find they have second-order accuracy. As a result, 
up to the second order, we find
\bea
  \frac{\phi_{a}(t,x+\alpha\Delta,y+\beta\Delta,z)}{\del z}
   &=& \frac{1}{2\Delta} \biggl[
     (1-\alpha)(1-\beta)\,
       \Bigl\{
         \phi_{a}(t,x,y,z+\Delta)-\phi_{a}(t,x,y,z-\Delta) 
       \Bigr\}
          \non \\
   && \quad \qquad
      + \alpha(1-\beta)\,
       \Bigl\{
         \phi_{a}(t,x+\Delta,y,z+\Delta)-\phi_{a}(t,x+\Delta,y,z-\Delta) 
       \Bigr\}
          \non \\
   && \quad \qquad
      + (1-\alpha)\beta\,
       \Bigl\{
         \phi_{a}(t,x,y+\Delta,z+\Delta)-\phi_{a}(t,x,y+\Delta,z-\Delta) 
       \Bigr\}
          \non \\
   && \quad \qquad
      + \alpha\beta\,
       \Bigl\{
         \phi_{a}(t,x+\Delta,y+\Delta,z+\Delta)
        -\phi_{a}(t,x+\Delta,y+\Delta,z-\Delta) 
       \Bigr\}
       \biggr]          \non \\
   && \quad \qquad
      + \CO(\Delta^2).
\eea

\newpage
\begin{table}
\caption{Five different sets of simulations.}
\label{tab:set1}
  \begin{center}
     \begin{tabular}{cccccc}
         Case & Lattice &
         Lattice spacing ($\Delta$) & $\zeta$ & Realization 
         & ${\rm Box~size}/H^{-1}$ \\
         & number & \protect{[}units of $t_{i}R(t)$\protect{]} &  &  & (at
         final time) \\
        \hline
        (a) & $128^3$ & $\sqrt{3}/8 $ & 10 & 100 & 1 (at 200) \\
        (b) & $256^3$ & $\sqrt{3}/16$ & 10 &  10 & 1 (at 200) \\
        (c) & $128^3$ & $\sqrt{3}/4 $ & 10 &  10 & 2 (at 200) \\
        (d) & $256^3$ & $\sqrt{3}/8 $ & 10 &  10 & 2 (at 200) \\
        (e) & $256^3$ & $\sqrt{3}/4 $ & 10 &  10 & 4 (at 200) \\
     \end{tabular}
  \end{center}
\end{table}

\begin{table}
\caption{Results of numerical simulations.}
\label{tab:set2}
  \begin{center}
     \begin{tabular}{ccccccc}
         Case & $\xi$ & $\la v \ra$ & $\la v^2 \ra$ & $\gamma$ 
         & $c$ & $\kappa$ \\
        \hline
        (a) & 0.72 & 0.63 & 0.50 & $2.0\pm0.2$ & $0.52\pm0.05$ 
            & $0.08\pm0.05$ \\
        (b) & 0.79 & 0.65 & 0.50 & $1.9\pm0.3$ & $0.48\pm0.04$
            & $0.08\pm0.04$ \\
        (c) & 0.77 & 0.60 & 0.50 & $1.8\pm0.2$ & $0.40\pm0.02$
            & $0.17\pm0.02$ \\
        (d) & 0.80 & 0.60 & 0.49 & $1.8\pm0.2$ & $0.42\pm0.03$
            & $0.16\pm0.03$ \\
        (e) & 0.80 & 0.60 & 0.50 & $1.8\pm0.2$ & $0.40\pm0.04$
            & $0.16\pm0.04$
     \end{tabular}
  \end{center}
\end{table}

\end{document}